\documentclass{iaus}
\usepackage{graphicx}

\def\la{\;
\raise0.3ex\hbox{$<$\kern-0.75em\raise-1.1ex\hbox{$\sim$}}\; }
\def\ga{\;
\raise0.3ex\hbox{$>$\kern-0.75em\raise-1.1ex\hbox{$\sim$}}\; }

\title[N-deficient and Fe-rich associated absorbers towards HE0141-3932] 
{Nitrogen-deficient and iron-rich associated absorbers with oversolar
metallicities towards the quasar HE0141-3932}

\author[Levshakov, Agafonova, Reimers, \etal ]   
{S. A. Levshakov$^1$, I. I. Agafonova$^1$, D. Reimers$^2$, C. Fechner$^2$, \break
E. Janknecht$^2$, \and S. Lopez$^3$}

\affiliation{$^1$A. F. Ioffe Physico-Technical Institute, 194021 St.Petersburg, Russia 
\break
$^2$Hamburger Sternwarte, Gojenbergsweg 112, 21029 Hamburg, Germany
\break 
$^3$Departamento de Astronomia, Universidad de Chile, Casilla 36-D, Santiago, Chile
}

\pubyear{2005}
\volume{228}  %% insert here the IAU Symposium No.
\pagerange{1--7}
\date{?? and in revised form ??}
\jname{From Lithium to Uranium: Elemental Tracers of Early Cosmic Evolution}
\editors{V. Hill, P. Fran\c{c}ois \& F. Primas, eds.}
\begin{document}

\maketitle

\begin{abstract}
HE0141-3932 ($z_{\rm em} = 1.80$)
is a bright blue radio-quite quasar which
reveals an emission line spectrum with an unusually weak Ly$\alpha$ line.
In addition, large redshift differences ($\Delta z = 0.05$) are observed
between high ionization and low ionization emission lines.
Absorption systems identified at 
$z_{\rm abs} = 1.78, 1.71$, and 1.68
show mild oversolar metallicities ($Z \approx 1-2Z_\odot$) and can be attributed
to the associated gas clouds ejected from the circumnuclear region.
The joint analysis of the emission and absorption lines leads to the
conclusion that this quasar is seen almost pole-on.
Its apparent luminosity may be Doppler boosted by $\sim 10$ times.
The absorbing gas shows high abundance of Fe, Mg, and Al
([Fe, Mg, Al/C] $\simeq 0.15\pm0.10$)
along with underabundance of N ([N/C] $\leq -0.5$).
This abundance pattern is at variance with current chemical evolution models
of QSOs predicting [N/C] $\ga 0$ and [Fe/C] $<0$
at $Z \sim Z_\odot$.
Full details of this work are given in  \cite{R05}.
\end{abstract}

%\firstsection % if your document starts with a section,
              % remove some space above using this command.
%\section{Introduction}

In the absorption spectrum of HE0141-3932 obtained with the VLT/UVES,
associated systems
are found at $z_{\rm abs} = 1.78, 1.71$, and 1.68. They
exhibit a wealth of metal absorptions including lines of Fe\,{\sc ii}.
Since these systems arise in gas
ejected from the circumnuclear region, careful analysis of the element
abundances can produce important clues about the enrichment
mechanisms and physical processes leading to the QSO/AGN phenomena.
The numbers presented in Table~1 are obtained with the Monte Carlo
Inversion (MCI) method~-- an advanced computational procedure designed
to estimate the physical parameters of optically thin
absorbers with varying gas density and turbulent velocity
field (\cite{L00}).

We conclude the following:
\begin{itemize}
\item
The analyzed absorption systems show high iron content,
[Fe/C] = $0.15\pm0.1$, [Fe/Mg] = $0.0\pm0.1$ ($z_{\rm abs}$=1.78),
but at the same time nitrogen
is strongly underabundant, [N/C] $\la -0.5$ ($z_{\rm abs}$=1.68).
These abundances are estimated in different associated systems
but they are representative for the bulk of circumnuclear gas since
the mass of the stellar population involved in the
enrichment of quasar's circumnuclear region is high,
$> 10^4 M_\odot$, (\cite{B03}) and, hence, large metallicity
gradients and sharp discontinuities due to enrichment by only a few stars
are unlikely.
\item
In the \cite{HF99} models of QSO's chemical evolution,
solar metallicity is reached after $\ga 0.2$ Gyr and is characterized by
a relative overabundance of nitrogen,
[N/C] $\ga 0$, and an underabundance of iron, [Fe/C] $< 0$.
Due to delay of 1~Gyr in Fe enrichment expected from
longer evolution of SNe~Ia which are the main source of iron,
the emission line ratios Fe\,{\sc ii}/C\,{\sc iv} and
Fe\,{\sc ii}/Mg\,{\sc ii} are proposed
to be a measure of QSO ages.
However, large values of [Fe/C] and [Fe/Mg]
are always associated in these models
with a considerable overabundance of nitrogen,
[N/C] $> 0.3$.
We do not observe such
relations in our systems and, hence, do not confirm this `iron clock'
model.
\item
The results obtained are
in line with \cite{MR01}, who showed that time scale
for enrichment by SNe~Ia is not unique but
a strong function of the adopted
stellar lifetimes, initial mass function, and star formation rate and can
vary by more than order of magnitude.
\end{itemize}

\begin{table}%\def~{\hphantom{0}} 
\begin{center}
\caption{Metal abundances at $z_{\rm abs} = 1.7817, 1.7103$, and 1.6838
towards  HE0141-3932 derived by the MCI procedure
(solar photospheric abundance for carbon is taken
from \cite{AP02}; for silicon, nitrogen and iron~--
from \cite{H01})
}
\label{tab1}
\begin{tabular}{lccccc} \hline
 &$z_{\rm abs}$=1.7817 & \multicolumn{2}{c}{$z_{\rm abs}$=1.7103} &
\multicolumn{2}{c}{$z_{\rm abs}$=1.6838}\\
Parameter &  & subsystem $A$ & subsystem $B$  & subsystem $A$
& subsystem $B$ \\
(1)$^a$ & (2) & (3)$^b$ & (4)$^b$ & (5) & (6) \\
\hline
$[Z_{\rm C}]$&$0.14\pm0.08$ &0.65 & 0.94 & $\la0.33$ & $0.08\pm0.10$\\
$[Z_{\rm N}]$&$<$0.15 & 0.56 & 0.58 &$\la-0.3$ & $-0.48\pm0.15$\\
$[Z_{\rm Mg}]$&$0.30\pm0.10$ & 0.6 & $\ldots$ &$\ldots$ & $\ldots$\\
$[Z_{\rm Al}]$&$0.30\pm0.10$ & 0.3 & $\ldots$ & $\ldots$ & $\ldots$\\
$[Z_{\rm Si}]$&$0.09\pm0.08$ & 0.33 & 0.94 & $\la$0.4 & $0.14\pm0.10$\\
$[Z_{\rm Fe}]$&$0.30\pm0.15$ & $\sim$1.5 & $\ldots$ & $\ldots$ & $\ldots$\\
\noalign{\smallskip}
$N$(H\,{\sc i}) &$(2.6\pm0.5)$E15 & 1.0E16 & $(2.6\pm0.5)$E14 &
$(1.1\pm0.2)$E14 & 2.5E15$^c$\\
$N$(C\,{\sc ii}) &$(1.64\pm0.08)$E13 & $(1.5\pm0.2)$E14 & $\la$1.0E12 &
$\la$1.6E11 & $(7.3\pm1.5)$E12 \\
$N$(C\,{\sc ii}$^\ast$) &$<$5.0E11 & $<$4.3E12 & $\ldots$ &
$\ldots$ & $<$1.5E12 \\
$N$(Mg\,{\sc ii}) &$(1.20\pm0.20)$E11 & $(5.3\pm1.5)$E12 & $\ldots$ &
$\ldots$ & $\ldots$ \\
$N$(Si\,{\sc ii}) &$(2.8\pm0.3)$E12 & $(1.2\pm0.1)$E13 & $\ldots$ &
 $\ldots$ & $\ldots$\\
$N$(Si\,{\sc ii}$^\ast$) &$<$1.3E11 & $<$2.0E11 & $\ldots$ &
$\ldots$& $\ldots$ \\
$N$(Fe\,{\sc ii}) &$(1.4\pm0.4)$E11 & $(5.5\pm2.0)$E11 &
$\ldots$ &$\ldots$&$\ldots$\\
$N$(Al\,{\sc iii})  &$(3.6\pm1.0)$E11 & $(1.4\pm0.3)$E12 & $\ldots$ &
$\ldots$ & $\ldots$\\
$N$(Si\,{\sc iii})  &7.5E12$^c$ & $(5.5\pm0.6)$E13 & $(2.3\pm1.2)$E11 &
$\la$8.5E10 & $(7.9\pm0.8)$E12\\
$N$(C\,{\sc iv})  &$(2.2\pm0.2)$E13 & $(1.7\pm0.2)$E15 & $(3.4\pm0.3)$E14 &
$(3.4\pm0.2)$E13 & $(2.3\pm0.1)$E14 \\
$N$(Si\,{\sc iv}) &$(4.5\pm0.4)$E12 &$(7.7\pm0.8)$E13 & $(1.6\pm0.5)$E12 &
$<$3.0E11 & $(1.2\pm0.1)$E13 \\
$N$(N\,{\sc v})  &$<$1.4E12 &$(1.8\pm0.2)$E14 & $(7.7\pm0.8)$E13 &
$(4.5\pm0.5)$E12 & $(1.3\pm0.1)$E13 \\ \hline
%\noalign{\bigskip}
\multicolumn{6}{l}{$^a$ $[Z_{\rm X}] = \log (N_{\rm X}/N_{\rm H}) -
\log (N_{\rm X}/N_{\rm H})_\odot$ ;\,\, $N$ is given in units of cm$^{-2}$;}\\
\multicolumn{6}{l}{$^b$\,\,estimated abundances are pure illustrative
since they are obtained under assumption of}\\
\multicolumn{6}{l}{the photoionization equilibrium whereas
the absorption system is non-equilibrial }\\
\multicolumn{6}{l}{(see \cite{R05} for details)}\\
\end{tabular}
\end{center}
\end{table}

\begin{acknowledgments}
S.A.L. thanks the IAU for the travel grant.
\end{acknowledgments}


\begin{thebibliography}{}


\bibitem[Allende Prieto \etal\ (2002)]{AP02}
{Allende Prieto, C., Lambert, D.L., \& Asplund, M.} 2002, 
\textit{ApJ} 573, L137

\bibitem[Baldwin \etal\ (2003)]{B03}
{Baldwin, J.A., Ferland, G.J., Korista, K.T., \etal } 2003, 
\textit{ApJ} 582, 590

\bibitem[Hamann \& Ferland (1999)]{HF99}
{Hamann, F., \& Ferland, G.} 1999, 
\textit{ARA\&A} 37, 487

\bibitem[Holweger (2001)]{H01}
{Holweger, H.} 2001, 
\textit{in Solar and Galactic Composition, ed. R. F.
Wimmer-Schweingruber, AIP Conf. Proc.} 598, 23

\bibitem[Levshakov \etal\ (2000)]{L00}
{Levshakov, S.A., Agafonova, I.I., \& Kegel, W.H.} 2000,
\textit{A\&A} 360, 833

\bibitem[Matteucci \& Recchi (2001)]{MR01}
{Matteucci, F., \& Recchi, S.} 2001, 
\textit{ApJ} 558, 351

\bibitem[Reimers \etal\ (2005)]{R05}
{Reimers, D., Janknecht, E., Fechner, C., Agafonova, I.I., Levshakov, S.A.,
\& Lopez, S.} 2005,
\textit{A\&A} 435, 17

\end{thebibliography}
\end{document}